\documentclass[twocolumn,showpacs,showkeys,amsmath,amssymb,superscriptaddress,floatfix,nofootinbib]{revtex4}

\usepackage{graphicx}
\usepackage{bm} 
\usepackage{subfigure}
	
\def\vec#1{\mathchoice{\mbox{\boldmath$\displaystyle#1$}}
{\mbox{\boldmath$\textstyle#1$}}
{\mbox{\boldmath$\scriptstyle#1$}}
{\mbox{\boldmath$\scriptscriptstyle#1$}}}
\makeatletter

\newcommand {\rootsNN} {\mbox{$\sqrt{s_{\rm NN}}$}}

\newcommand {\pt} {\mbox{$p_{\rm T}$}~}
\newcommand {\mt} {\mbox{$m_{\rm T}$}~}

\newcommand {\dnde} { \mbox{$\left< dN_{ch}/d\eta \right>^{1/3}$}~}
\newcommand {\dndef} { \mbox{$\left< dN_{ch}/d\eta \right>$}}

\begin{document}
 
\title{Pion-kaon femtoscopy in Pb--Pb collisions at \rootsNN=2.76
  TeV modeled in (3+1)D hydrodynamics coupled to Therminator 2 and the
  effect of delayed kaon emission 
}

\author{Adam Kisiel} 
\email{kisiel@if.pw.edu.pl}
\affiliation{Faculty of Physics, Warsaw University of Technology, 
ul. Koszykowa 75, PL-00662, Warsaw, Poland}

\begin{abstract}
Non-identical particle femtoscopy measures the size of the system
emitting particles (``radius'') in heavy-ion collisions as well as the
difference 
between mean emission space-time coordinates of two particle
species (``emission asymmetry''). The system created in such
collisions at the LHC behaves collectively and its dynamics is well 
described by hydrodynamic models.
A significant emission asymmetry between pions and kaons, coming from
collective flow, enhanced by contribution from flowing resonances is
predicted. We present calculations within the (3+1)D viscous
hydrodynamic model coupled to statistical hadronization code
Therminator 2, corresponding to Pb--Pb collisions at
\rootsNN=2.76~TeV. We obtain femtoscopic radii and emission asymmetry 
for pion-kaon pairs as a function of collision centrality.
The radii grow linearly with cube root of particle multiplicity density.
The emission asymmetry is negative
and comparable to the radius, indicating that pions are emitted
closer to the center of the system and/or later than kaons.
Recent ALICE Collaboration measurements of identical kaon femtoscopy
shows that kaons are emitted, on average, 2.1~fm/$c$ later than
pions. We modify our calculation by introducing such delay and find
that the system source size is only weakly affected. In contrast the
pion-kaon emission asymmetry is directly sensitive to such delays and
the modified calculation shows significantly lower values of
asymmetry. Therefore we propose the measurement of the pion-kaon
femtoscopic correlation function as a sensitive probe of the time
delays in particle emission.

\end{abstract}



\maketitle 


\section{Introduction} 
\label{sec:intro} 

The system created in collisions of heavy-ions at ultra-relativistic
energies is dynamically expanding and cooling. In the early stages of
the evolution it is thought to be in a deconfined phase (the
Quark-Gluon Plasma), where the matter behaves as a strongly
coupled liquid with small specific
viscosity~\cite{Adams:2005dq,Adcox:2004mh,Back:2004je,Arsene:2004fa}. Models
which employ hydrodynamic equations to describe this behavior are
successful in reproducing many of the observables in such
collisions. The most common are the transverse momentum spectra and
elliptic flow, which are also modified by event-by-event
fluctuations. This behavior of the system in momentum space is driven
by space-time characteristics of the source -- its size and gradients
in pressure. Therefore, the correct description of the momentum space
observables must be accompanied by a proper simulation of its
space-time structure and its dynamics. The basic principles of such
description have been established at
RHIC~\cite{Broniowski:2008vp,Pratt:2009hu}. They have then been
applied at the
LHC~\cite{Kisiel:2014upa,Bozek:2017kxo,Karpenko:2012yf,Karpenko:2010te,Shapoval:2014wya}
to describe the data on identical particle
femtoscopy~\cite{Aamodt:2011mr,Adam:2015vja,Adam:2015vna,Acharya:2017qtq}. They
include significant pre-thermal flows, an Equation of State that does
not include the first-order phase transition, a careful treatment of
strongly-decaying resonances as well as possible addition of viscosity.

The femtoscopic radii obtained from identical pion analysis agree well
with model calculations~\cite{Adam:2015vna}, indicating that the
dynamics of the system at the LHC is well described. Similar data for
kaons show novel features~\cite{Acharya:2017qtq}. The femtoscopic
radii for kaons are larger than expected 
from the naive ``\mt~scaling'' argument. The radii are also
larger than predicted by Therminator 2 model, which includes
hydrodynamic evolution of the system followed by the statistical
hadronization. A second calculation presented
in~\cite{Acharya:2017qtq} is based on hydrodynamical model coupled to
the hadronic rescattering code~\cite{Shapoval:2014wya} and is able to
reproduce the larger values of the kaon radii. The increase is
attributed to the delay of the emission time of kaons, coming from the
rescattering via the $K^{*}$ meson. Experimental analysis with the
theoretical formalism proposed in~\cite{Shapoval:2014wya} shows that
on average kaons are emitted later than pions by 2.1~fm/$c$. This
result is then interpreted as evidence for the extended
``rescattering'' phase in the evolution of the heavy-ion
collision. Confirmation of the existence of such phase has profound
consequences for modeling such collisions and experimental data 
interpretation. 

Femtoscopic technique is not limited to pairs of identical
particles. For pairs of non-identical particles, the correlation
arises from then the Final State 
Interactions (FSI), that is Coulomb when both particles in the pair
are charged and Strong interaction when both particles are
hadrons. The original motivation for the formulation of the
non-identical particle femtoscopy formalism was to measure the
difference in average time of the emission of various 
types of particles~\cite{Lednicky:1995vk}, which was called ``emission
asymmetry''. It was later realized that spatial emission asymmetry
will produce equivalent asymmetry
signal~\cite{Lednicky:2001qv,Lednicky:2005tb}.  Such spatial asymmetry
arises naturally in a hydrodynamically expanding system, where thermal
and flow velocities are comparable. Heavy-ion collisions produce such
a system. In a detailed analysis of the non-identical particle
correlations at RHIC energies~\cite{Kisiel:2009eh} the emission
asymmetry between pions, kaons, and protons was studied in
detail. It was found that the emission asymmetry coming from the
radial flow in the system is enhanced, for pion-kaon and pion-proton
pairs, by additional non-trivial effects coming from the decay of
flowing resonances. A complete set of emission asymmetries for three
pair types was presented. Also the formalism was introduced and its
validity was tested. The results show that the same formalism should
be applicable in heavy-ion collisions at LHC.

This work contains two main results. It presents the calculation of
the pion-kaon femtoscopic correlation functions for Pb--Pb collisions
at the LHC energies, for selected centralities. The calculation is
carried out within the model consisting of (3+1)D viscous
hydrodynamics coupled to Therminator 2 statistical hadronization,
resonance propagation and decay code. System size and emission
asymmetry between pions and kaons is extracted as a function of
centrality. These results can be directly compared to experimental
results. In particular experimental acceptance constraints similar to
those imposed by the ALICE detector are applied. In a separate
calculation the model is modified to include the additional 
emission time delay for kaons, as suggested by the experimental data
from ALICE~\cite{Acharya:2017qtq}. Several values of the time delay
are explored, including the $2.1$~fm/$c$ value obtained
experimentally. We explore how the introduction of time delay
influences the extracted emission asymmetry and system size. We
present separate sets of emission asymmetries, as a function of
collision centrality, for selected values of time delay. Experimental
data from ALICE can be directly compared to these datasets. Most
importantly the comparison to data may reveal if the emission
asymmetry deduced from identical kaon femtoscopy is also observed in
pion-kaon emission asymmetry. Such observation would be an important
and independent confirmation of the existence of the ``rescattering''
phase of the Pb--Pb collision at the LHC.  

The work is organized as follows. In Sec.~\ref{sec:model} the model
used in this work is described and the datasample which was analyzed
is characterized. In Sec.~\ref{sec:nonidform} the formalism of the 
non-identical particle correlations is briefly
introduced. Sec.~\ref{sec:results} discusses the main results of this
work -- the pion-kaon femtoscopic correlation functions as well as
system sizes and emission asymmetries which were extracted from their
analysis.  

\section{(3+1)D Hydro and Therminator 2 models}
\label{sec:model}

The model used in this work is composed of two parts. The collective
expansion is modeled in the (3+1)D viscous hydrodynamics. The details
of the implementation and the formalism of the model is presented
in~\cite{Bozek:2011ua,Bozek:2012qs,Bozek:2014hwa}. The particle
emission is implemented in the statistical hadronization and resonance
propagation and decay simulation code Therminator 2~\cite{Chojnacki:2011hb}.  

In particular the calculations presented in this work have been
intentionally performed on exactly the same generated model dataset,
which was used in our previous work on identical particle
femtoscopy~\cite{Kisiel:2014upa}. These calculations were later used by
the ALICE Collaboration for comparison with experimental data for
identical kaon femtoscopy~\cite{Acharya:2017qtq}. The discrepancies
between our model calculation and data are an explicit scientific
motivation for the studies in this work. We refer the reader to the
works mentioned above for details of the model calculations. Here we
only briefly mention the important features.

We use the viscous hydrodynamic model, following the second order
Israel-Stewart  equations \cite{Gale:2012rq}. Hard equation of state
is used~\cite{Broniowski:2008vp,Pratt:2009hu}, in particular a
parametrization interpolating between lattice QCD
results~\cite{Borsanyi:2010cj} at high temperatures and the hadron gas
equation of state at low temperatures. All chemical potentials are set
to zero. We use smooth initial conditions for the hydrodynamic 
evolution, given by the Glauber model. The initial time for the
hydrodynamic evolution is $0.6$~fm/c, viscosity coefficients are
$\eta/s=0.08$ and $\zeta/s=0.04$,  and the  freeze-out temperature
$T_t=140$~MeV.

The calculation is performed for five sets of initial conditions,
corresponding to impact parameter $b$ values (in fm) for the Pb--Pb
collisions at the \rootsNN=2.76 TeV: 3.1, 5.7, 7.4, 8.7, and
9.9~fm. They correspond, in terms of the average particle multiplicity
density \dndef, to given centrality ranges at the
LHC~\cite{Aamodt:2010cz}: 0--10\%, 10--20\%, 20--30\%, 30--40\%, and
40--50\%. 

Therminator 2~\cite{Chojnacki:2011hb} code then performs a statistical
hadronization on the freeze-out hypersurfaces obtained from the hydro
model via the Cooper-Frye formalism. Chemical and kinetic freeze-outs
are equated. Importantly the model does not include hadronic
rescattering. It does however implement the propagation and decay (in
cascades if necessary) of all known hadronic resonances. The final
output from the model is a set of events, each composed of final-state
particles, with information on the particle identity, momentum and
space-time freeze-out coordinates provided for each of them. 

\section{Non-identical particle femtoscopy formalism}
\label{sec:nonidform}

The formalism for non-identical particle femtoscopy is described in
great detail in~\cite{Lednicky:2005tb,Kisiel:2009eh}. Here we only
briefly introduce the main concepts.

The femtoscopic correlation function measures the conditional
probability to measure two particles of a given type at a certain
relative momentum $\vec{q}$. In order to eliminate the trivial
dependence on particle acceptance, such probability is normalized to
the product of probabilities to measure each particle
separately. Experimentally in heavy-ion collisions the measurement
consists of constructing the 
distribution of pairs of particles of given types $X$ and $Y$ coming
from the same event and storing their relative momenta in the
distribution $A_{XY}(\vec{q})$. Then similar procedure is repeated,
but the two particles come from two different events, giving the
reference distribution $B_{XY}(\vec{q})$. The correlation function is
then: 
\begin{equation}
  C(\vec{q}) = A_{XY}(\vec{q})/B_{XY}(\vec{q})
  \label{eq:CfromQ}
\end{equation}
For a pair consisting of a charged pion and a charged kaon, the
femtoscopic correlation arises from the Strong and Coulomb Final State
Interaction. Currently no model of heavy-ion collisions implements
such interactions, therefore it must be introduced `a posteriori' via
the so-called ``afterburner'' procedure~\cite{Kisiel:2009eh}. The
pairs of particles are obtained from model events, and samples
$A_{XY}$ and $B_{XY}$ are constructed in a procedure resembling the
experimental one as closely as possible. However for each model pair
going into the $A_{XY}$ sample an additional weight corresponding to
the square of the module of the Bethe-Salpeter amplitude $\Psi_{XY}$
of the pair is added~\cite{Lednicky:2005tb}. The correlation function
is then a ratio of the weighted $A_{XY}$ sample to the $B_{XY}$
sample. The amplitude $\Psi_{XY}$ will be described in detail later in
the manuscript. 

The model procedure described above produces a correlation function
with femtoscopic effects as well as all the other event-wide
correlations present in the model. Some additional non-femtoscopic
correlations were studied in the previous
works~\cite{Kisiel:2017gip,Graczykowski:2014tsa} and were found to be
significant for pion-kaon pairs. However, it was also shown that such
correlation can be efficiently corrected for with a data-driven
procedure. Since this work focuses on the analysis of the femtoscopic
effect, we do not study these additional correlations. Instead we
employ a modified procedure, where the $B_{XY}$ sample is simply the
$A_{XY}$ sample without femtoscopic weights. In such calculation the
non-femtoscopic correlations are not
present~\cite{Kisiel:2009eh,Kisiel:2017gip}.

The theoretical interpretation of the correlation function assumes
that it is expressed as: 
\begin{equation}
C(\vec{k^{*}}) = {{\int S(\bf{r^{*}}, \vec{k^{*}})
  |\Psi_{XY}(\bf{r^{*}}, \vec{k^{*}})|^{2}} \over {\int
  S(\bf{r^{*}}, \vec{k^{*}})}} 
\label{eq:cfrompsi}
\end{equation}
where $\bf{r^{*}}={\bf x}_{1}-{\bf x}_{2}$ is a relative space-time
separation of the two particles at the moment of their
creation. $\vec{k^{*}}$ is the momentum of the first particle in the 
Pair Rest Frame, so it is half of the pair relative momentum in this
frame (for identical particles $\vec{q} = 2\vec{k^{*}}$). $S$ is the
source emission function and can be interpreted as a probability to
emit a given particle pair from a given set of emission points with
given momenta.

For a charged pion-charged kaon pair $\Psi_{\pi K}$ contains
contributions from the Strong and 
Coulomb interaction~\cite{Lednicky:2005tb}. However for this
particular pair, the Strong interaction is expected to be small. The
femtoscopic signal is dominated by the Coulomb interaction, especially
for the emission asymmetry signature. Therefore in this work we use a
simplification: we only consider the Coulomb part of the
interaction. We use it self-consistently first to calculate the model
correlation functions and later in the fitting procedure to extract
the the system size and emission asymmetry. With this modification
we have:
\begin{equation}
\Psi_{XY} = \sqrt{A_C(\eta)} \left [ e^{-ik^{*}r^{*}}F(-i\eta, 1,
    i\xi) \right ],
\label{eq:CoulFun}
\end{equation}
where $A_C$ is the Gamov factor, $\xi = k^{*}r^{*}(1+\cos\theta^{*})$,
$\eta = 1/(k^{*}a_C)$, and $F$ is the confluent hypergeometric
function. $\theta^{*}$ is the angle between $\vec{k^{*}}$ and
$\vec{r^{*}}$ and $a_C$ is the Bohr radius which is equal to $\pm 248.52$~fm
for the pion-kaon pair. The correlation function then shows a positive
correlation effect for unlike-sign pion-kaon pairs, and a negative
correlation effect for like-sign pairs. This $\Psi_{\pi K}$ is used as
a basis to calculate the weight for the model correlation function
calculation and in the fitting procedure.

\begin{figure}[tb]
\begin{center}
\includegraphics[angle=0,width=0.45 \textwidth]{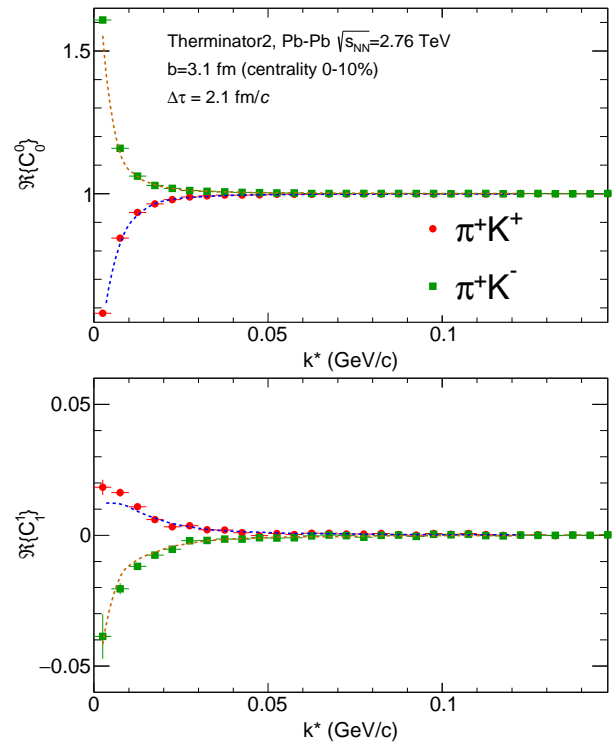}
\end{center}
\vspace{-6.5mm}
\caption{(Color online) Spherical Harmonics components of the example
  charged pion-charged kaon femtoscopic 
  correlation function calculated in the (3+1)D Hydrodynamic model
  coupled to Therminator 2 code for central Pb--Pb collisions
  at $\sqrt{s_{\rm NN}}=2.76$~TeV. Lines show the fit to the
  correlation function (see text for details). 
\label{fig:pikcorrfuncs}}
\end{figure}

This ``afterburner'' procedure employing weights, as described above,
is used to calculate the femtoscopic correlation functions
for all charge combinations of charged pion-charged kaon pairs. The
functions are stored in the Spherical Harmonics
representation~\cite{Kisiel:2009iw}. We only analyze the two
components of this representation, the $l=0,m=0$ 
and the real part of the $l=1,m=1$. It was shown
in~\cite{Kisiel:2009eh} that these two components contain the
relevant signals for the system size and emission asymmetry. The pions
and kaons were selected in the \pt range of $0.15$ to $2.5$~GeV/$c$
and pseudorapidity range $|\eta|<1.0$, which corresponds to the
reconstruction and PID acceptance of the ALICE detector.
Two example correlation functions, one for like-sign another for
unlike-sign pion-kaon pair, are shown in
Fig.~\ref{fig:pikcorrfuncs}. A positive(negative) correlation effect,
coming from the Coulomb attraction(repulsion) for
unlike-sign(like-sign) pion-kaon can be clearly seen. Similarly the
$\Re{C_1^1}$ clearly deviates from zero, indicating a non-zero
emission asymmetry between pions and kaons.

\subsection{Correlation function fitting}
\label{sec:fitting}

The model correlation function is then analyzed in a procedure closely
resembling an experimental one. First it is assumed that the source is
an ellipsoid with a Gaussian density profile. It has different widths
in three directions defined in the Longitudinally Co-Moving System
(LCMS), where the total longitudinal momentum of the pair
vanishes. The directions are: ``long'' along the beam axis, ``out''
along the pair transverse momentum and ``side'' perpendicular to the
other two. From the symmetry of the heavy-ion collision it is expected
that the emission asymmetry between particles arises only in the
``out''
direction~\cite{Lednicky:2001qv,Lednicky:2005tb,Kisiel:2009eh}. Its
magnitude is another model parameter. The final form of the assumed
emission function is then:
\begin{equation}
S(\vec{r}) \approx \exp\left(-\frac{[r_{out} -
    \mu_{out}]^{2}}{2\sigma_{out}^{2}}
  -\frac{r_{side}^{2}}{2\sigma_{side}^{2}}
  -\frac{r_{long}^{2}}{2\sigma_{long}^{2}}
   \right),
\label{eq:Slcms}
\end{equation}
where $\sigma$ are the sizes of the system in the three directions and
$\mu$ is the emission asymmetry. The correlation function for
non-identical particles is only weakly sensitive to the details of the
three-dimensional shape of $S$. Such details were much more precisely
studied through identical particle
femtoscopy~\cite{Aamodt:2011mr,Adam:2015vja,Adam:2015vna,Acharya:2017qtq}. We
limit the number of independent fitting parameters by
fixing $\sigma_{side} = \sigma_{out}$ and $\sigma_{long} = 1.3
\sigma_{out}$. The values of the scaling coefficients are based on
corresponding values of system sizes from identical pion
femtoscopy~\cite{Adam:2015vna}.  
In  this work we  focus instead on the emission asymmetry, which is
not accessible in that technique. As a result we have 
only two independent fit parameters: $\sigma_{out}$ characterizing the
overall system size as well as $\mu_{out}$ containing information of
the pion-kaon emission asymmetry. 

\begin{figure}[tb]
\begin{center}
\includegraphics[angle=0,width=0.45 \textwidth]{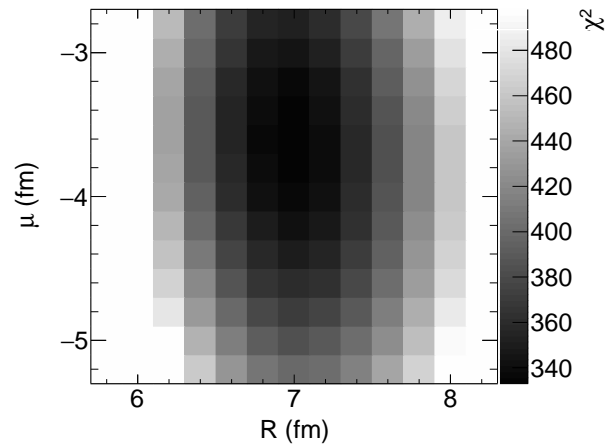}
\end{center}
\vspace{-6.5mm}
\caption{Example of the $\chi^2$ values obtained in the fitting
  procedure for one of the pion-kaon femtoscopic correlation functions
  (see text for details).  
\label{fig:pikchi2map}}
\end{figure}

For a given set of $\sigma_{out}$ and $\mu_{out}$ values a ``fit''
correlation function is calculated according to Eq.~\eqref{eq:cfrompsi}
with $\Psi$ given by Eq.~\eqref{eq:CoulFun}. Then a $\chi^2$ value is
calculated between this function a the ``experimental-like'' one
calculated for Therminator 2 data. The calculation is repeated for all
combinations of $\sigma_{out}$ and $\mu_{out}$ values in pre-defined
ranges. An example result of such calculation is shown in
Fig.~\ref{fig:pikchi2map}. The minimization procedure is then
employed to find the $\sigma_{out}$ and $\mu_{out}$ values that
minimize the $\chi^2$ value. This set is the result of the fit. The
procedure is implemented in the CorrFit
software~\cite{Kisiel:2004fcn}. The fitting procedure also accounts
for the so-called ``purity'' of the sample, or the percentage of the
pairs that form the Gaussian core of the system. The values for this
purity parameter depend mostly on the percentage of pions and kaons
that come from strongly decaying resonances. Their abundances
depend on the temperature of the chemical freeze-out. This temperature
is very similar for RHIC and LHC calculations in Therminator,
therefore in this work we have used the values estimated in the
previous work for RHIC~\cite{Kisiel:2009eh}. The $\chi^2$ landscape
shown in Fig.~\ref{fig:pikchi2map} also reveals very small correlation
between the $\sigma_{out}$ and $\mu_{out}$ fitting results. This has
been observed consistently for all performed fits. In indicates that
uncertainties of $\sigma_{out}$ and $\mu_{out}$ are uncorrelated.

\section{Results}
\label{sec:results}

We calculate the femtoscopic correlation for all four charge
combination of the pion-kaon pair: $\pi^{+}K^{+}$, $\pi^{+}K^{-}$,
$\pi^{-}K^{+}$, and $\pi^{-}K^{-}$. All correlations are fitted
independently. At the end the result is averaged between all four
charge combinations. The calculation has been performed for the event
samples obtained from the (3+1)D hydrodynamic code coupled to
Therminator 2 statistical hadronization, resonance propagation and
decay code. The five samples used correspond to Pb--Pb collisions at
the selected centralities. On the figures data for the five samples
are plotted at the corresponding \dnde.

\begin{figure}[tb]
\begin{center}
\includegraphics[angle=0,width=0.45 \textwidth]{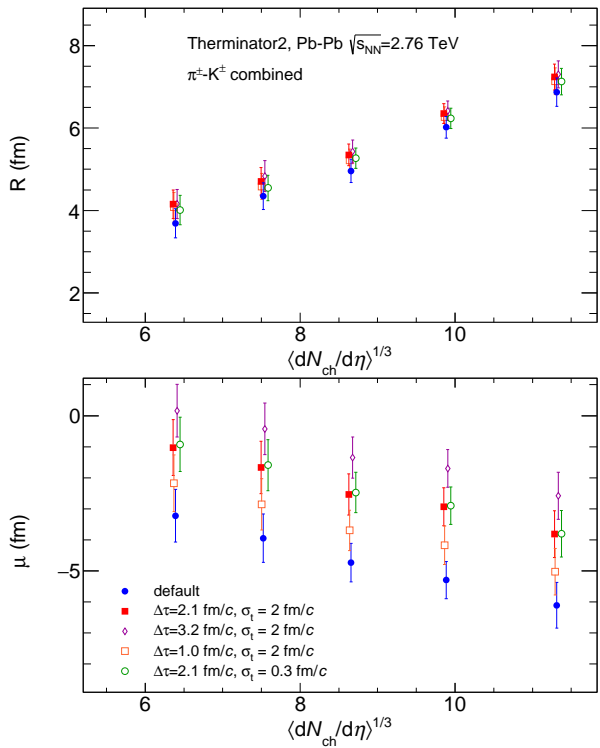}
\end{center}
\vspace{-6.5mm}
\caption{(Color online) Source size (upper panel) and pion-kaon
  emission asymmetry (lower panel) from pion-kaon correlation
  functions calculated in the Therminator 2 model for Pb--Pb collisions
  at $\sqrt{s_{\rm NN}}=2.76$~TeV for selected centralities. Blue
  open points show default calculation. Red closed points, orange open
  squares, green closed squares and violet open diamonds show
  calculations with selected values of additional time delay for
  kaons (see text for details). Some points were shifted slightly in
  the $x$-direction for clarity. 
\label{fig:piksigmutd}}
\end{figure}

The standard calculation is performed on the generated events
directly. Following the experimental insight~\cite{Acharya:2017qtq} we
also performed the calculations with a specific modification. For each
kaon its emission time is modified by adding a delay $\Delta\tau$,
distributed according to a Gaussian, with a certain width and
mean. Firstly a calculation with a mean time delay of $2.1$~fm/$c$ and a
width of $2$~fm/$c$ was performed. Then three other calculations were
done, one with the mean changed to $1$~fm/$c$, next with the mean changed
to $3.2$~fm/$c$ and the last one with the width changed to $0.3$~fm/$c$. The
results of the fits to all the calculated correlation functions are
presented in Fig.~\ref{fig:piksigmutd}.

The figure shows the set of predictions for the pion-kaon source size
and asymmetry in heavy-ion collisions at the LHC energy. The system
size grows with event multiplicity, the dependence is to a good
approximation linear. This is expected and understood, as similar
increase has been consistently observed in all measurements for
identical pion femtoscopy. The pion-kaon system size is a convolution
of the size of the system emitting pions and kaons at a given
velocity. Therefore, it is mostly influenced by the source radius
which is larger, which usually is the one for pions.

The emission asymmetry in the default calculation is universally
negative. This means that pions are emitted closer to the center of
the system and/or later than kaons. This emission asymmetry is
relatively large, comparable to the system size. It was shown
in~\cite{Kisiel:2009eh} that it is coming from the spatial asymmetry
produced by a flow of primordial pions and kaons. This asymmetry is
further enhanced by the qualitative difference in the way that
resonance decays influence the emission pattern of pions and
kaons. The pion decay momentum for most common resonances is similar
or larger than a pion mass. The direction of the velocity of such pion
is heavily 
randomized, and is no longer strongly influenced by the flow field. As
a result the average emission point of pions from resonances is close
to the geometrical center of the source. In contrast for kaons the
decay momentum is usually small compared to the kaon mass. The parent
resonance is usually quite heavy, therefore it is strongly pushed by
the flow field. After decay the kaon velocity direction is still
strongly correlated with the parent one. Therefore the average
emission point for kaons is strongly pushed by the flow to the edge of
the source, producing a large emission asymmetry with respect to
pions.

When a time delay is introduced for kaons, the fit results are visibly
changed. The overall system size is only slightly affected. It grows
by approximately $0.5$~fm for all calculations. The increase is larger
when the introduced time delay is larger. The width of the time delay
distribution has a smaller but still visible effect on size. The
calculation with time delay distribution width of $0.3$~fm/$c$ gives a
size about $0.1-0.2$~fm smaller than the calculation with the time
delay width of $2$~fm/$c$.

\begin{figure}[tb]
\begin{center}
\includegraphics[angle=0,width=0.45 \textwidth]{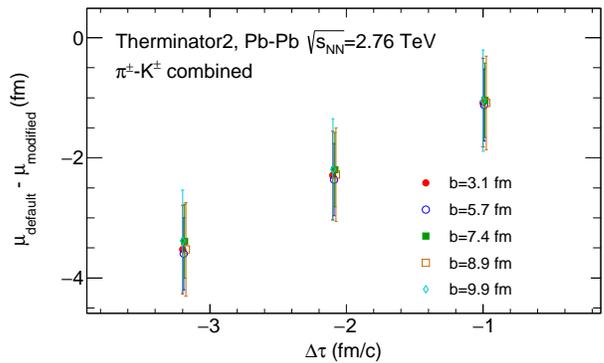}
\end{center}
\vspace{-6.5mm}
\caption{(Color online) The difference in the extracted pion-kaon
  emission asymmetry between the default calculation for the
  Therminator 2 model for Pb--Pb collisions at $\sqrt{s_{\rm
      NN}}=2.76$~TeV and the calculations with additional kaon
  emission time delay, plotted as a function of the value of
  additional shift. Some points were shifted slightly in the
  $x$-direction for clarity. 
\label{fig:pikdelaydep}}
\end{figure}

The kaon emission time delay has a direct and strong effect on the
pion-kaon emission asymmetry. As expected the delay significantly
decreases the emission asymmetry. When the time delay is increased to
$3.2$~fm the emission asymmetry even turns positive for most
peripheral collisions. The value of introduced time delay is shifting
the extracted emission asymmetry and the value of this shift is
independent of centrality (and therefore system size).

In Fig.~\ref{fig:pikdelaydep} the difference in the extracted emission
asymmetry between the default calculation and the calculations with
the kaon emission time delay is plotted as a function of the value of
this introduced delay. The 
dependence between the two seems to be a direct one-to-one
correspondence, independent of the system size. In other words this
calculation shows that pion-kaon emission asymmetry is directly and
linearly sensitive to any delays in emission time of kaons. The
calculation shows, that the experimental measurement of the pion-kaon
emission asymmetry can be a very sensitive cross-check of the model
interpretation of identical pion and kaon femtoscopy data given
in~\cite{Acharya:2017qtq}. In particular if the interpretation given
in this work is correct, than ALICE should observe pion-kaon emission
asymmetry of approximately $-4$~fm in most central collisions, instead
of $-6$~fm predicted by the default Therminator 2 calculation.

It should be noted that the analysis of the kaon emission time
presented in~\cite{Acharya:2017qtq} are given for central events
only. It is postulated that the emission delay is a result of
rescattering via the $K^{*}$ resonance. If that is the case (and given
that the chemical freeze-out temperature and consequently the relative
abundance of this resonance changes little with centrality), then the
value of the delay should be similar at other centralities too. This
work shows the predictions for them as well. On the other hand if the
experimentally observed asymmetry will be different at other
centralities, the calculations shown in Fig.~\ref{fig:piksigmutd} can
be used to estimate the kaon emission time delay from the data.

\section{Summary}
\label{sec:summary}

We have presented the first calculations of pion-kaon femtoscopic
correlation function for Pb--Pb collisions at the \rootsNN=2.76 TeV at
selected collision centralities. System size and emission asymmetry
was extracted for each pion-kaon charge combination and collision
centrality separately. The extracted system size is observed to
linearly increase with cube root of the charged particle multiplicity
density. The emission asymmetry is large and negative, indicating that
pions are emitted closer to the center of the source and/or later than
kaons. Such asymmetry is naturally expected in a hydrodynamically
flowing medium, where large fraction of particles is produced via
resonance decay. The results are also qualitatively consistent with the
calculations at top RHIC collision energies.

Following the experimental results for identical kaon femtoscopy the
calculation has been modified by introducing an emission time delay
for kaons. The pion-kaon asymmetry is shown to be directly and
linearly sensitive to such delay. The introduction of a time delay of
$2.1$~fm reduced the pion-kaon asymmetry to approximately $-4$~fm for
central collisions, compared to approximately $-6$~fm for default
calculation. Such difference should be measurable in the ALICE
experiment. The experimental data on pion-kaon emission asymmetry can
be a direct and sensitive test of the existence of emission time delay
for kaons. The confirmation of its existence would be a strong and
independent argument for the importance of the hadronic rescattering
phase at the LHC. 

\section*{Acknowledgment}
This work has been supported by the Polish National Science Centre
under grant No. UMO-2014/13/B/ST2/04054.

\bibliography{citations}

\begin{thebibliography}{30}
\expandafter\ifx\csname natexlab\endcsname\relax\def\natexlab#1{#1}\fi
\expandafter\ifx\csname bibnamefont\endcsname\relax
  \def\bibnamefont#1{#1}\fi
\expandafter\ifx\csname bibfnamefont\endcsname\relax
  \def\bibfnamefont#1{#1}\fi
\expandafter\ifx\csname citenamefont\endcsname\relax
  \def\citenamefont#1{#1}\fi
\expandafter\ifx\csname url\endcsname\relax
  \def\url#1{\texttt{#1}}\fi
\expandafter\ifx\csname urlprefix\endcsname\relax\def\urlprefix{URL }\fi
\providecommand{\bibinfo}[2]{#2}
\providecommand{\eprint}[2][]{\url{#2}}

\bibitem[{\citenamefont{Adams et~al.}(2005)}]{Adams:2005dq}
\bibinfo{author}{\bibfnamefont{J.}~\bibnamefont{Adams}} \bibnamefont{et~al.}
  (\bibinfo{collaboration}{STAR}), \bibinfo{journal}{Nucl. Phys.}
  \textbf{\bibinfo{volume}{A757}}, \bibinfo{pages}{102} (\bibinfo{year}{2005}).

\bibitem[{\citenamefont{Adcox et~al.}(2005)}]{Adcox:2004mh}
\bibinfo{author}{\bibfnamefont{K.}~\bibnamefont{Adcox}} \bibnamefont{et~al.}
  (\bibinfo{collaboration}{PHENIX}), \bibinfo{journal}{Nucl. Phys.}
  \textbf{\bibinfo{volume}{A757}}, \bibinfo{pages}{184} (\bibinfo{year}{2005}).

\bibitem[{\citenamefont{Back et~al.}(2005)}]{Back:2004je}
\bibinfo{author}{\bibfnamefont{B.~B.} \bibnamefont{Back}} \bibnamefont{et~al.},
  \bibinfo{journal}{Nucl. Phys.} \textbf{\bibinfo{volume}{A757}},
  \bibinfo{pages}{28} (\bibinfo{year}{2005}).

\bibitem[{\citenamefont{Arsene et~al.}(2005)}]{Arsene:2004fa}
\bibinfo{author}{\bibfnamefont{I.}~\bibnamefont{Arsene}} \bibnamefont{et~al.}
  (\bibinfo{collaboration}{BRAHMS}), \bibinfo{journal}{Nucl. Phys.}
  \textbf{\bibinfo{volume}{A757}}, \bibinfo{pages}{1} (\bibinfo{year}{2005}).

\bibitem[{\citenamefont{Broniowski et~al.}(2008)\citenamefont{Broniowski,
  Chojnacki, Florkowski, and Kisiel}}]{Broniowski:2008vp}
\bibinfo{author}{\bibfnamefont{W.}~\bibnamefont{Broniowski}},
  \bibinfo{author}{\bibfnamefont{M.}~\bibnamefont{Chojnacki}},
  \bibinfo{author}{\bibfnamefont{W.}~\bibnamefont{Florkowski}},
  \bibnamefont{and} \bibinfo{author}{\bibfnamefont{A.}~\bibnamefont{Kisiel}},
  \bibinfo{journal}{Phys.Rev.Lett.} \textbf{\bibinfo{volume}{101}},
  \bibinfo{pages}{022301} (\bibinfo{year}{2008}).

\bibitem[{\citenamefont{Pratt}(2009)}]{Pratt:2009hu}
\bibinfo{author}{\bibfnamefont{S.}~\bibnamefont{Pratt}},
  \bibinfo{journal}{Nucl. Phys.} \textbf{\bibinfo{volume}{A830}},
  \bibinfo{pages}{51c} (\bibinfo{year}{2009}).

\bibitem[{\citenamefont{Kisiel et~al.}(2014)\citenamefont{Kisiel, Gałażyn,
  and Bożek}}]{Kisiel:2014upa}
\bibinfo{author}{\bibfnamefont{A.}~\bibnamefont{Kisiel}},
  \bibinfo{author}{\bibfnamefont{M.}~\bibnamefont{Gałażyn}},
  \bibnamefont{and} \bibinfo{author}{\bibfnamefont{P.}~\bibnamefont{Bożek}},
  \bibinfo{journal}{Phys. Rev.} \textbf{\bibinfo{volume}{C90}},
  \bibinfo{pages}{064914} (\bibinfo{year}{2014}), \eprint{1409.4571}.

\bibitem[{\citenamefont{Bożek}(2017)}]{Bozek:2017kxo}
\bibinfo{author}{\bibfnamefont{P.}~\bibnamefont{Bożek}},
  \bibinfo{journal}{Phys. Rev.} \textbf{\bibinfo{volume}{C95}},
  \bibinfo{pages}{054909} (\bibinfo{year}{2017}), \eprint{1702.01319}.

\bibitem[{\citenamefont{Karpenko et~al.}(2013)\citenamefont{Karpenko, Sinyukov,
  and Werner}}]{Karpenko:2012yf}
\bibinfo{author}{\bibfnamefont{I.}~\bibnamefont{Karpenko}},
  \bibinfo{author}{\bibfnamefont{Y.}~\bibnamefont{Sinyukov}}, \bibnamefont{and}
  \bibinfo{author}{\bibfnamefont{K.}~\bibnamefont{Werner}},
  \bibinfo{journal}{Phys.Rev.} \textbf{\bibinfo{volume}{C87}},
  \bibinfo{pages}{024914} (\bibinfo{year}{2013}), \eprint{1204.5351}.

\bibitem[{\citenamefont{Karpenko and Sinyukov}(2010)}]{Karpenko:2010te}
\bibinfo{author}{\bibfnamefont{I.~A.} \bibnamefont{Karpenko}} \bibnamefont{and}
  \bibinfo{author}{\bibfnamefont{{\relax Yu}.~M.} \bibnamefont{Sinyukov}},
  \bibinfo{journal}{Phys. Rev.} \textbf{\bibinfo{volume}{C81}},
  \bibinfo{pages}{054903} (\bibinfo{year}{2010}), \eprint{1004.1565}.

\bibitem[{\citenamefont{Shapoval et~al.}(2014)\citenamefont{Shapoval,
  Braun-Munzinger, Karpenko, and Sinyukov}}]{Shapoval:2014wya}
\bibinfo{author}{\bibfnamefont{V.~M.} \bibnamefont{Shapoval}},
  \bibinfo{author}{\bibfnamefont{P.}~\bibnamefont{Braun-Munzinger}},
  \bibinfo{author}{\bibfnamefont{I.~A.} \bibnamefont{Karpenko}},
  \bibnamefont{and} \bibinfo{author}{\bibfnamefont{{\relax Yu}.~M.}
  \bibnamefont{Sinyukov}}, \bibinfo{journal}{Nucl. Phys.}
  \textbf{\bibinfo{volume}{A929}}, \bibinfo{pages}{1} (\bibinfo{year}{2014}),
  \eprint{1404.4501}.

\bibitem[{\citenamefont{Aamodt et~al.}(2011{\natexlab{a}})}]{Aamodt:2011mr}
\bibinfo{author}{\bibfnamefont{K.}~\bibnamefont{Aamodt}} \bibnamefont{et~al.}
  (\bibinfo{collaboration}{ALICE Collaboration}), \bibinfo{journal}{Phys.Lett.}
  \textbf{\bibinfo{volume}{B696}}, \bibinfo{pages}{328}
  (\bibinfo{year}{2011}{\natexlab{a}}).

\bibitem[{\citenamefont{Adam et~al.}(2015)}]{Adam:2015vja}
\bibinfo{author}{\bibfnamefont{J.}~\bibnamefont{Adam}} \bibnamefont{et~al.}
  (\bibinfo{collaboration}{ALICE}), \bibinfo{journal}{Phys. Rev.}
  \textbf{\bibinfo{volume}{C92}}, \bibinfo{pages}{054908}
  (\bibinfo{year}{2015}), \eprint{1506.07884}.

\bibitem[{\citenamefont{Adam et~al.}(2016)}]{Adam:2015vna}
\bibinfo{author}{\bibfnamefont{J.}~\bibnamefont{Adam}} \bibnamefont{et~al.}
  (\bibinfo{collaboration}{ALICE}), \bibinfo{journal}{Phys. Rev.}
  \textbf{\bibinfo{volume}{C93}}, \bibinfo{pages}{024905}
  (\bibinfo{year}{2016}), \eprint{1507.06842}.

\bibitem[{\citenamefont{Acharya et~al.}(2017)}]{Acharya:2017qtq}
\bibinfo{author}{\bibfnamefont{S.}~\bibnamefont{Acharya}} \bibnamefont{et~al.}
  (\bibinfo{collaboration}{ALICE}), \bibinfo{journal}{Phys. Rev.}
  \textbf{\bibinfo{volume}{C96}}, \bibinfo{pages}{064613}
  (\bibinfo{year}{2017}), \eprint{1709.01731}.

\bibitem[{\citenamefont{Lednicky et~al.}(1996)\citenamefont{Lednicky,
  Lyuboshits, Erazmus, and Nouais}}]{Lednicky:1995vk}
\bibinfo{author}{\bibfnamefont{R.}~\bibnamefont{Lednicky}},
  \bibinfo{author}{\bibfnamefont{V.~L.} \bibnamefont{Lyuboshits}},
  \bibinfo{author}{\bibfnamefont{B.}~\bibnamefont{Erazmus}}, \bibnamefont{and}
  \bibinfo{author}{\bibfnamefont{D.}~\bibnamefont{Nouais}},
  \bibinfo{journal}{Phys. Lett.} \textbf{\bibinfo{volume}{B373}},
  \bibinfo{pages}{30} (\bibinfo{year}{1996}).

\bibitem[{\citenamefont{Lednicky}(2001)}]{Lednicky:2001qv}
\bibinfo{author}{\bibfnamefont{R.}~\bibnamefont{Lednicky}}, in
  \emph{\bibinfo{booktitle}{{International Workshop on the Physics of the Quark
  Gluon Plasma Palaiseau, France, September 4-7, 2001}}}
  (\bibinfo{year}{2001}), \eprint{nucl-th/0112011}.

\bibitem[{\citenamefont{Lednicky}(2009)}]{Lednicky:2005tb}
\bibinfo{author}{\bibfnamefont{R.}~\bibnamefont{Lednicky}},
  \bibinfo{journal}{Phys. Part. Nucl.} \textbf{\bibinfo{volume}{40}},
  \bibinfo{pages}{307} (\bibinfo{year}{2009}), \eprint{nucl-th/0501065}.

\bibitem[{\citenamefont{Kisiel}(2010)}]{Kisiel:2009eh}
\bibinfo{author}{\bibfnamefont{A.}~\bibnamefont{Kisiel}},
  \bibinfo{journal}{Phys. Rev.} \textbf{\bibinfo{volume}{C81}},
  \bibinfo{pages}{064906} (\bibinfo{year}{2010}), \eprint{0909.5349}.

\bibitem[{\citenamefont{Bozek}(2012)}]{Bozek:2011ua}
\bibinfo{author}{\bibfnamefont{P.}~\bibnamefont{Bozek}},
  \bibinfo{journal}{Phys.Rev.} \textbf{\bibinfo{volume}{C85}},
  \bibinfo{pages}{034901} (\bibinfo{year}{2012}).

\bibitem[{\citenamefont{Bo\.zek and Wyskiel-Piekarska}(2012)}]{Bozek:2012qs}
\bibinfo{author}{\bibfnamefont{P.}~\bibnamefont{Bo\.zek}} \bibnamefont{and}
  \bibinfo{author}{\bibfnamefont{I.}~\bibnamefont{Wyskiel-Piekarska}},
  \bibinfo{journal}{Phys. Rev.} \textbf{\bibinfo{volume}{C85}},
  \bibinfo{pages}{064915} (\bibinfo{year}{2012}).

\bibitem[{\citenamefont{Bo\.zek}(2014)}]{Bozek:2014hwa}
\bibinfo{author}{\bibfnamefont{P.}~\bibnamefont{Bo\.zek}},
  \bibinfo{journal}{Phys. Rev.} \textbf{\bibinfo{volume}{C89}},
  \bibinfo{pages}{044904} (\bibinfo{year}{2014}).

\bibitem[{\citenamefont{Chojnacki et~al.}(2012)\citenamefont{Chojnacki, Kisiel,
  Florkowski, and Broniowski}}]{Chojnacki:2011hb}
\bibinfo{author}{\bibfnamefont{M.}~\bibnamefont{Chojnacki}},
  \bibinfo{author}{\bibfnamefont{A.}~\bibnamefont{Kisiel}},
  \bibinfo{author}{\bibfnamefont{W.}~\bibnamefont{Florkowski}},
  \bibnamefont{and}
  \bibinfo{author}{\bibfnamefont{W.}~\bibnamefont{Broniowski}},
  \bibinfo{journal}{Comput.Phys.Commun.} \textbf{\bibinfo{volume}{183}},
  \bibinfo{pages}{746} (\bibinfo{year}{2012}).

\bibitem[{\citenamefont{Gale et~al.}(2013)\citenamefont{Gale, Jeon, Schenke,
  Tribedy, and Venugopalan}}]{Gale:2012rq}
\bibinfo{author}{\bibfnamefont{C.}~\bibnamefont{Gale}},
  \bibinfo{author}{\bibfnamefont{S.}~\bibnamefont{Jeon}},
  \bibinfo{author}{\bibfnamefont{B.}~\bibnamefont{Schenke}},
  \bibinfo{author}{\bibfnamefont{P.}~\bibnamefont{Tribedy}}, \bibnamefont{and}
  \bibinfo{author}{\bibfnamefont{R.}~\bibnamefont{Venugopalan}},
  \bibinfo{journal}{Phys. Rev. Lett.} \textbf{\bibinfo{volume}{110}},
  \bibinfo{pages}{012302} (\bibinfo{year}{2013}).

\bibitem[{\citenamefont{Borsanyi et~al.}(2010)}]{Borsanyi:2010cj}
\bibinfo{author}{\bibfnamefont{S.}~\bibnamefont{Borsanyi}}
  \bibnamefont{et~al.}, \bibinfo{journal}{JHEP} \textbf{\bibinfo{volume}{11}},
  \bibinfo{pages}{077} (\bibinfo{year}{2010}).

\bibitem[{\citenamefont{Aamodt et~al.}(2011{\natexlab{b}})}]{Aamodt:2010cz}
\bibinfo{author}{\bibfnamefont{K.}~\bibnamefont{Aamodt}} \bibnamefont{et~al.}
  (\bibinfo{collaboration}{ALICE Collaboration}),
  \bibinfo{journal}{Phys.Rev.Lett.} \textbf{\bibinfo{volume}{106}},
  \bibinfo{pages}{032301} (\bibinfo{year}{2011}{\natexlab{b}}).

\bibitem[{\citenamefont{Kisiel}(2017)}]{Kisiel:2017gip}
\bibinfo{author}{\bibfnamefont{A.}~\bibnamefont{Kisiel}},
  \bibinfo{journal}{Acta Phys. Polon.} \textbf{\bibinfo{volume}{B48}},
  \bibinfo{pages}{717} (\bibinfo{year}{2017}).

\bibitem[{\citenamefont{Graczykowski et~al.}(2014)\citenamefont{Graczykowski,
  Kisiel, Janik, and Karczmarczyk}}]{Graczykowski:2014tsa}
\bibinfo{author}{\bibfnamefont{Å.~K.} \bibnamefont{Graczykowski}},
  \bibinfo{author}{\bibfnamefont{A.}~\bibnamefont{Kisiel}},
  \bibinfo{author}{\bibfnamefont{M.~A.} \bibnamefont{Janik}}, \bibnamefont{and}
  \bibinfo{author}{\bibfnamefont{P.}~\bibnamefont{Karczmarczyk}},
  \bibinfo{journal}{Acta Phys. Polon.} \textbf{\bibinfo{volume}{B45}},
  \bibinfo{pages}{1993} (\bibinfo{year}{2014}), \eprint{1409.8120}.

\bibitem[{\citenamefont{Kisiel and Brown}(2009)}]{Kisiel:2009iw}
\bibinfo{author}{\bibfnamefont{A.}~\bibnamefont{Kisiel}} \bibnamefont{and}
  \bibinfo{author}{\bibfnamefont{D.~A.} \bibnamefont{Brown}},
  \bibinfo{journal}{Phys. Rev.} \textbf{\bibinfo{volume}{C80}},
  \bibinfo{pages}{064911} (\bibinfo{year}{2009}).

\bibitem[{\citenamefont{Kisiel}(2004)}]{Kisiel:2004fcn}
\bibinfo{author}{\bibfnamefont{A.}~\bibnamefont{Kisiel}},
  \bibinfo{journal}{Nukleonika} \textbf{\bibinfo{volume}{49}},
  \bibinfo{pages}{s81} (\bibinfo{year}{2004}).

\end{thebibliography}

\end{document}